# Explicit Circular Harmonic Inversions of Exponential Radon Transform[1]

**Jiangsheng You**[0], **Geyang Du**[1], **Gengsheng L Zeng**[2] and **Zhengrong Liang**[3]

Cubic Imaging LLC[0], 18 Windemere Dr., Andover, MA 01810
Department of Mathematics[1], Peking University, Beijing 100871
Department of Radiology[2], University of Utah, Salt Lake City, UT 84108
Department of Radiology[3], State University of New York, Stony Brook, NY 11794

jshyou@gmail.com
gydu@pku.edu.cn
larry@ucair.med.utah.edu
jerome.liang@stonybrookmedicine.edu

## Abstract

Using Plemelj formula we obtain three circular harmonic inversion formulas of the exponential Radon transform with complex coefficients. We also derive two different range conditions and prove that Novikov's range condition does imply the traditional range condition for real coefficients.

## I.   Introduction and Preliminary

We denote by $R^2$ the two-dimensional plane with the point representations of $\vec{r} = (x, y)$ in the Cartesian coordinates and $\vec{r} = (r, \varphi)$ in the polar coordinates, respectively. We frequently use these notations: $\vec{\theta} = (\cos\theta, \sin\theta)$, $\vec{\theta}^{\perp} = (-\sin\theta, \cos\theta)$, $I = (-1, 1)$, $D = \{\vec{r} : |\vec{r}| < 1\}$, $S^1 = \{\vec{r} : |\vec{r}| = 1\}$, $\bar{I}$ and $\bar{D}$ for the closure of the corresponding open sets, and $C$ for the complex plane. For $\mu \in C$, the exponential Radon transform (ERT) of $f(x, y)$ is defined by

$$p(s, \theta, \mu) = \int_{-\infty}^{\infty} f(s\vec{\theta} + t\vec{\theta}^{\perp}) e^{\mu t} dt . \qquad (1.0)$$

The evenness condition of the ERT is expressed as

$$p(-s, \theta, \mu) = p(s, \theta + \pi, -\mu) . \qquad (1.1)$$

Without specifically mentioning throughout this paper, we use assumption (**A1**) stated as

(**A1**)    *$f(x, y)$ is a smooth complex function with support in $\bar{D}$, $\mu$ is a complex constant and its ERT $p(s, \theta, \mu)$ is a smooth function with support in $\bar{I} \times S^1$.*

The ERT has many practical applications as discussed in [1-4] and references therein. One of the most important applications is the single photon emission computed tomography (SPECT) [1, 2], in which $f(x, y)$ is the density distribution of the radioactivity tracer injected in the human

---

body tissue, $\mu$ represents the linear attenuation of human body tissue to Gamma rays and $p(s,\theta,\mu)$ is obtained through modifying the counts of emitted photons reaching to the detector. The goal of SPECT is to estimate $f(x,y)$ from measurement $p(s,\theta,\mu)$. All the parameters in SPECT take non-negative real values. However, in some other applications such as nuclear magnetic resonance (NMR) imaging, ultrasonic tomography and Doppler tomography described in [3, 4], $\mu$ may take pure imaginary values. Nonetheless, the main task in these applications is to find $f(x,y)$ from measurement $p(s,\theta,\mu)$ provided that $\mu$ is known. Mathematically, this task is equivalent to inverting (1.0), which is often called image reconstruction in tomographic imaging. The ERT can be regarded as a linear operator from functions of $\overline{D}$ to functions of $\overline{I} \times S^1$. The range condition is to describe a set of relations under which a function $p(s,\theta,\mu)$ is the ERT of certain $f(x,y)$.

The study of the explicit inversion and range conditions of the ERT has been a research topic for many years from both mathematical interests and practical needs. A substantial review on the known results and open issues can be found in [5, 6] and many review papers from journals such as *Inverse problems*, *IEEE Transactions on Medical Imaging*, and *Physics in Medicine and Biology*. The fundamental work on the explicit inversion formulas and range conditions for the ERT are from [7-10]. Recent works [11-14] provide a new way to study this topic, for example solving transport equation is the key technique in Novikov's work. Inspired by Cormack's historical works [15, 16], in this paper, we continue to explore the circular harmonic approach to study the explicit inversion and range conditions of the ERT. The circular harmonics concern the Fourier series expansion of a function with respect to the angular variable. In the expression of the Fourier series expansion, $f(r,\varphi)$ and $p(s,\theta,\mu)$ can be written as

$$f(r,\varphi) = \sum_{-\infty}^{\infty} f_n(r)e^{in\varphi}, \tag{1.2}$$

$$p(s,\theta,\mu) = \sum_{-\infty}^{\infty} p_n(s,\mu)e^{in\theta}. \tag{1.3}$$

Throughout this paper $p_n(s,\mu)$ always stands for the circular harmonics defined in (1.3). The goal of the circular harmonic approach is to establish explicit relations between $f_n(r)$ and $p_n(s,\mu)$.

The circular harmonic approach has been investigated in [7-9, 15-20] and many other papers. The relation derived in [7, 8] is through the Fourier transform of $p_n(s,\mu)$ with respect to $s$. An explicit formula was obtained in [9] for the case of $f(x,y) = f(r)$. The algorithm in [17] first used the relation in [7, 8] to obtain the circular harmonics of the Fourier transform of $f(x,y)$ and then found $f_n(r)$ through inverse Fourier transform. The numerical method in [18] is the direct application of the Fourier series expansion to the convolution kernel without using an explicit formula. In [19], Puro derived two types of closed-form inversion formulas for pure imaginary coefficients. The main idea of [19] is to use complex analysis to calculate the singular integrals. Later on Puro and Garin extended work [19] to the case of complex axially attenuation coefficients in [20]. More work on this subject can be found in [21-24].

In this paper, for the complex $\mu$, we use Plemelj formula to calculate the singular integrals to derive the inversion formulas. The first type of inversion is the explicit and stable circular harmonic formulas without Chebyshev polynomials by using two equivalent inversion formulas from [4]. The second type of inversion includes two formulas derived by the range condition and orthogonal relations of the exponential Chebyshev functions. We also analyze the relations among several range conditions. For $\mu \in R^1$, we show that Novikov's range condition yields the traditional range condition in [8, 10].

In Cormack's works [15, 16], the Chebyshev polynomials play a very important role in establishing the relations between $f_n(r)$ and $p_n(s,0)$. For $n \geq 0$ and $s \in R^1$, let $T_n(s)$ and $U_n(s)$ stand for the Chebyshev polynomials of the first and second kind, respectively. The detailed orthogonal relations and numerical properties of the Chebyshev polynomials can be found in [25]. To study the ERT, we introduce the exponential Chebyshev functions previously suggested in [19]. For $s, r \in R^1, r \neq 0$, integer $n$ and $\mu \in C$, the exponential Chebyshev functions $\tilde{T}_n(s,r,\mu)$ and $\tilde{U}_{n-1}(s,r,\mu)$ are defined as

$$\tilde{T}_n(s,r,\mu) = \frac{1}{2}[e^{i\mu\sqrt{s^2-r^2}}(\frac{s+\sqrt{s^2-r^2}}{r})^n + e^{-i\mu\sqrt{s^2-r^2}}(\frac{s-\sqrt{s^2-r^2}}{r})^n] \tag{1.4}$$

$$\tilde{U}_{n-1}(s,r,\mu) = \frac{r}{2\sqrt{s^2-r^2}}[e^{i\mu\sqrt{s^2-r^2}}(\frac{s+\sqrt{s^2-r^2}}{r})^n - e^{-i\mu\sqrt{s^2-r^2}}(\frac{s-\sqrt{s^2-r^2}}{r})^n] \tag{1.5}$$

Notice that $\tilde{T}_n(s,r,0)$ and $\tilde{U}_n(s,r,0)$ become the conventional Chebyshev polynomials $T_n(\frac{s}{r})$ and $U_n(\frac{s}{r})$, respectively. For $\tilde{T}_n(s,r,\mu)$ and $\tilde{U}_n(s,r,\mu)$, we collect several simple relations

$$\tilde{T}_n(-s,r,\mu) = (-1)^n \tilde{T}_n(s,r,-\mu), \ \tilde{T}_n(s,r,-\mu) = \tilde{T}_{-n}(s,r,\mu), \tag{1.6}$$

$$\tilde{U}_n(-s,r,\mu) = (-1)^n \tilde{U}_n(s,r,-\mu), \ \tilde{U}_{n-1}(s,r,-\mu) = -\tilde{U}_{-n-1}(s,r,\mu), \tag{1.7}$$

$$\tilde{U}_{n-1}(s,r,\mu) \pm \frac{r}{\sqrt{s^2-r^2}}\tilde{T}_n(s,r,\mu) = \pm\frac{r}{\sqrt{s^2-r^2}}e^{\pm i\mu\sqrt{s^2-r^2}}(\frac{s\pm\sqrt{s^2-r^2}}{r})^n. \tag{1.8}$$

Throughout this paper, for integer $n$ we will use the following definition

$$\text{sgn}(n) = \begin{cases} 1 & n \geq 0 \\ -1 & n < -1 \end{cases}$$

With aforementioned notations, we state several known facts from [5, 17, 19] as Proposition 1 with proofs. The moment condition is extended to an exterior expression.

**Proposition 1**. *Under assumption* (**A1**), $f_n(r)$ *and* $p_n(s,\mu)$ *satisfy*

$$p_n(s,\mu) = 2\int_{|s|}^{\infty} \frac{\tilde{T}_n(s,r,\mu)}{\sqrt{r^2-s^2}} f_n(r) r dr, \tag{1.9}$$

$$p_n(-s,\mu) = (-1)^n p_n(s,-\mu), \tag{1.10}$$

$$\int_{|s| \geq r} s^m e^{-i\text{sgn}(n)\mu s} p_n(s,\mu) ds = 0, \ r \geq 0, \ 0 \leq m < |n|, \tag{1.11}$$

$$\int_{-\infty}^{\infty} s^m e^{-i\text{sgn}(n)\mu s} p_n'(s,\mu) ds = 0, \ 0 \leq m < |n|. \tag{1.12}$$

**Proof**. Both (1.9) and (1.10) have been derived in the early literature, for completeness we provide the proof in this paper. For $r = 0$, (1.11) was previously derived in [5, 19]. Here we prove that the range condition (1.11) holds for $r > 0$. The derivation of (1.12) is straightforward and is necessary to derive the inversion formulas in next section.

Calculating the Fourier series of (1.0), we obtain (1.9) as follows.

$$\int_0^{2\pi} p(s,\theta,\mu)e^{-in\theta}d\theta = \int_0^{2\pi}[\int_{-\infty}^{\infty} f(s\vec{\theta}+t\vec{\theta}^{\perp})e^{\mu t}dt]e^{-in\theta}d\theta$$

$$= \int_0^{\infty} e^{\mu t}[\int_0^{2\pi} f(s\vec{\theta}+t\vec{\theta}^{\perp})e^{-in\theta}d\theta]dt + \int_{-\infty}^0 e^{\mu t}[\int_0^{2\pi} f(s\vec{\theta}+t\vec{\theta}^{\perp})e^{-in\theta}d\theta]dt \tag{1.13}$$

Let $r=\sqrt{t^2+s^2}$ , $t=\pm\sqrt{r^2-s^2}$ , $\varphi=\arccos(s/r)$ and $f(s\vec{\theta}\pm\sqrt{r^2-s^2}\ \vec{\theta}^{\perp})=f(r,\theta\pm\varphi)$ , then

$$\int_0^{\infty} e^{\mu t}[\int_0^{2\pi} f(s\vec{\theta}+t\vec{\theta}^{\perp})e^{-in\theta}d\theta]dt = \int_0^{\infty} e^{\mu\sqrt{r^2-s^2}}[\int_0^{2\pi} f(s\vec{\theta}+\sqrt{r^2-s^2}\ \vec{\theta}^{\perp})e^{-in\theta}d\theta]d\sqrt{r^2-s^2}$$

$$= \int_0^{\infty} e^{\mu\sqrt{r^2-s^2}}e^{in\varphi}[\int_0^{2\pi} f(r,\theta+\varphi)e^{-in(\theta+\varphi)}d\theta]\frac{r}{\sqrt{r^2-s^2}}dr$$

$$= \int_{|s|}^{\infty} e^{\mu\sqrt{r^2-s^2}}e^{in\varphi}\frac{f_n(r)r}{\sqrt{r^2-s^2}}dr \tag{1.14}$$

$$= \int_{|s|}^{\infty} e^{\mu\sqrt{r^2-s^2}}\left[\frac{s+i\sqrt{r^2-s^2}}{r}\right]^n\frac{f_n(r)r}{\sqrt{r^2-s^2}}dr$$

$$\int_{-\infty}^0 e^{\mu t}[\int_0^{2\pi} f(s\vec{\theta}+t\vec{\theta}^{\perp})e^{-in\theta}d\theta]dt = \int_{-\infty}^0 e^{-\mu\sqrt{r^2-s^2}}[\int_0^{2\pi} f(s\vec{\theta}-\sqrt{r^2-s^2}\ \vec{\theta}^{\perp})e^{-in\theta}d\theta]d[-\sqrt{r^2-s^2}\ ]$$

$$= \int_{|s|}^{\infty} e^{-\mu\sqrt{r^2-s^2}}e^{-in\varphi}[\int_0^{2\pi} f(r,\theta-\varphi)e^{-in(\theta-\varphi)}d\theta]\frac{r}{\sqrt{r^2-s^2}}dr$$

$$= \int_{|s|}^{\infty} e^{-\mu\sqrt{r^2-s^2}}e^{-in\varphi}\frac{f_n(r)r}{\sqrt{r^2-s^2}}dr \tag{1.15}$$

$$= \int_{|s|}^{\infty} e^{-\mu\sqrt{r^2-s^2}}\left[\frac{s-i\sqrt{r^2-s^2}}{r}\right]^n\frac{f_n(r)r}{\sqrt{r^2-s^2}}dr$$

Combine identities from (1.13) to (1.15) we have

$$p_n(s,\mu) = \int_0^{2\pi} p(s,\theta,\mu)e^{-in\theta}d\theta$$

$$= \int_{|s|}^{\infty}\left\{e^{\mu\sqrt{r^2-s^2}}\left[\frac{s+i\sqrt{r^2-s^2}}{r}\right]^n + e^{-\mu\sqrt{r^2-s^2}}\left[\frac{s-i\sqrt{r^2-s^2}}{r}\right]^n\right\}\frac{f_n(r)r}{\sqrt{r^2-s^2}}dr$$

$$= \int_{|s|}^{\infty}\left\{e^{i\mu\sqrt{s^2-r^2}}\left[\frac{s+\sqrt{s^2-r^2}}{r}\right]^n + e^{-i\mu\sqrt{s^2-r^2}}\left[\frac{s-\sqrt{s^2-r^2}}{r}\right]^n\right\}\frac{f_n(r)r}{\sqrt{r^2-s^2}}dr \tag{1.16}$$

$$= 2\int_{|s|}^{\infty}\frac{\widetilde{T}_n(s,r,\mu)f_n(r)r}{\sqrt{r^2-s^2}}dr$$

Relation (1.10) is straightforward from (1.6) and (1.9).

Let $\rho=\sqrt{s^2+t^2}$ , $\phi=\mathrm{acos}(s/\rho)$ , it follows that $s=\rho\cos\phi$ and $t=\rho\sin\phi$ . We rewrite the left hand side of (1.11) as

$$\int\limits_{|s|\geq r} s^m e^{-i\operatorname{sgn}(n)\mu s}\, p_n(s,\mu)\,ds$$

$$= \int\limits_0^{2\pi} e^{-in\theta}\,d\theta\,[\int\limits_{|s|\geq r}\int\limits_{-\infty}^{\infty} f(s\vec{\theta}+t\vec{\theta}^{\perp})s^m e^{\mu\theta-i\operatorname{sgn}(n)\mu s}\,dt\,ds]$$

$$= \int\limits_0^{2\pi} e^{-in\theta}\,d\theta\,[\int\limits_r^{\infty}\int\limits_0^{2\pi} f(\rho,\theta+\phi)(\rho\cos\phi)^m e^{\mu\rho\sin\phi-i\operatorname{sgn}(n)\mu\rho\cos\phi}\,\rho\,d\rho\,d\phi]$$

$$= \int\limits_r^{\infty}\int\limits_0^{2\pi}[\int\limits_0^{2\pi} f(\rho,\theta+\phi)e^{-in(\theta+\phi)}\,d\theta]e^{in\phi}(\rho\cos\phi)^m e^{\mu\rho[\sin\phi-i\operatorname{sgn}(n)\cos\phi]}\,\rho\,d\rho\,d\phi \qquad (1.17)$$

$$= \int\limits_r^{\infty}\int\limits_0^{2\pi} f_n(\rho)e^{in\phi}(\rho\cos\phi)^m e^{-i\mu\rho[\operatorname{sgn}(n)\cos\phi+i\sin\phi]}\,\rho\,d\rho\,d\phi$$

$$= \frac{1}{2^m}\int\limits_r^{\infty}[\int\limits_0^{2\pi} e^{in\phi}(e^{i\phi}+e^{-i\phi})^m e^{-i\operatorname{sgn}(n)\mu\rho e^{i\operatorname{sgn}(n)\phi}}\,d\phi]f_n(\rho)\rho^{m+1}\,d\rho$$

Let $z=e^{i\phi}$, we have the series expansion

$$e^{in\phi}(e^{i\phi}+e^{-i\phi})^m e^{-i\operatorname{sgn}(n)\mu\rho e^{i\operatorname{sgn}(n)\phi}} = z^n(z+z^{-1})^m\sum_{k\geq 0}\frac{1}{k!}(-i\operatorname{sgn}(n)\mu\rho)^k z^{\operatorname{sgn}(n)k}\ . \qquad (1.18)$$

For $0\leq m\lhd|n|$, (1.18) only contains either negative or positive powers of $z$, thus

$$\int\limits_0^{2\pi} e^{in\phi}(e^{i\phi}+e^{-i\phi})^m e^{-i\operatorname{sgn}(n)\mu\rho e^{i\operatorname{sgn}(n)\phi}}\,d\phi = 0\ . \qquad (1.19)$$

Combining preceding relations from (1.17) to (1.19), we obtain (1.11).

For $p_n'(s,\mu)$, $0\leq m\lhd|n|$, using the integration by parts, we have

$$\int\limits_{-\infty}^{\infty} s^m e^{-i\operatorname{sgn}(n)\mu s}p_n'(s,\mu)\,ds = -\int\limits_{-\infty}^{\infty}[ms^{m-1}-i\operatorname{sgn}(n)\mu s^m]e^{-i\operatorname{sgn}(n)\mu s}p_n(s,\mu)\,ds = 0\ . \qquad (1.20)$$

This completes the proof. □

# II.   Explicit Inversion Formulas of (1.0)

We will use the complex analysis method to evaluate several singular integrals in deriving explicit inversion formulas of (1.0). For the general knowledge on the complex analysis, we refer to [26]. We also use the Plemelj formula to evaluate the Hilbert transform on the unit circle, we refer to [27] for more details on the Plemelj formula. The main result in this section is the closed-form formulas to express $f_n(r)$ as an integral operator of $p_n'(s,\mu)$. We also obtain an exterior inversion formula and a new range condition.

**Theorem 1**. *Under assumption* (**A1**)*, let $p_n'(s,\mu)$ be the derivative of $p_n(s,\mu)$ with respect to $s$, then $f_n(r)$ can be obtained by the following formula*

$$f_n(r) = \frac{1}{2\pi r}\int\limits_{-\infty}^{\infty}\operatorname{sgn}(n)\widetilde{U}_{n-1}(s,r,-\mu)p_n'(s,\mu)\,ds - \frac{1}{2\pi r}\int\limits_{|s|>r}\frac{\operatorname{sgn}(s)r}{\sqrt{s^2-r^2}}\widetilde{T}_n(s,r,-\mu)p_n'(s,\mu)\,ds \qquad (2.1)$$

**Proof**. Equation (2.1) closely mimics Cormack's inversion formula for the Radon transform. We cite two types of inversion formulas from Theorem 2 and Corollary of [5] as follows

$$f(r,\varphi) = \frac{1}{4\pi^2}\int\limits_0^{2\pi} e^{-\mu\vec{r}\cdot\vec{\theta}^{\perp}}[\frac{\partial}{\partial l}\int\limits_{-\infty}^{\infty}\frac{e^{-i\mu(l-s)}}{l-s}p(\theta,s,\mu)\,ds]|_{l=\vec{r}\cdot\vec{\theta}}\,d\theta\ . \qquad (2.2)$$

$$f(r,\varphi) = \frac{1}{4\pi^2}\int\limits_0^{2\pi} e^{-\mu\vec{r}\cdot\vec{\theta}^{\perp}}[\frac{\partial}{\partial l}\int\limits_{-\infty}^{\infty}\frac{e^{i\mu(l-s)}}{l-s}p(\theta,s,\mu)\,ds]|_{l=\vec{r}\cdot\vec{\theta}}\,d\theta\ . \qquad (2.3)$$

Rewrite (2.2) to

$$f(r,\varphi) = \frac{1}{4\pi^2} \int_0^{2\pi} e^{\mu r \sin(\theta-\varphi)} \left[ \int_{-\infty}^{\infty} \frac{e^{-i\mu(r\cos(\theta-\varphi)-s)}}{r\cos(\theta-\varphi)-s} p'(\theta,s,\mu) ds \right] d\theta$$

$$= \frac{1}{4\pi^2} \int_0^{2\pi} e^{\mu r \sin(\theta-\varphi)} \left[ \int_{-\infty}^{\infty} \frac{e^{-i\mu(r\cos(\theta-\varphi)-s)}}{r\cos(\theta-\varphi)-s} \sum_{-\infty}^{\infty} p_n'(s,\mu) e^{in\theta} ds \right] d\theta \qquad (2.4)$$

$$= \frac{1}{4\pi^2} \sum_{-\infty}^{\infty} e^{in\varphi} \int_{-\infty}^{\infty} \left[ \int_0^{2\pi} \frac{e^{\mu r \sin\theta - i\mu(r\cos\theta-s)}}{r\cos\theta-s} e^{in\theta} d\theta \right] p_n'(s,\mu) ds.$$

Then (2.4) becomes

$$f(r,\varphi) = \sum_{-\infty}^{\infty} e^{in\varphi} \frac{1}{\pi} \int_{-\infty}^{\infty} \left[ \frac{1}{2\pi} \int_0^{2\pi} \frac{e^{i\mu(s-re^{i\theta})}}{r(e^{i\theta}+e^{-i\theta})-2s} e^{in\theta} d\theta \right] p_n'(s,\mu) ds. \qquad (2.5)$$

Similarly, we rewrite (2.3) as

$$f(r,\varphi) = \sum_{-\infty}^{\infty} e^{in\varphi} \frac{1}{\pi} \int_{-\infty}^{\infty} \left[ \frac{1}{2\pi} \int_0^{2\pi} \frac{e^{-i\mu(s-re^{-i\theta})}}{r(e^{i\theta}+e^{-i\theta})-2s} e^{in\theta} d\theta \right] p_n'(s,\mu) ds. \qquad (2.6)$$

Define two functions

$$K_n^+(r,s,\mu) = \frac{1}{2\pi} \int_0^{2\pi} \frac{e^{i\mu(s-re^{i\theta})}}{r(e^{i\theta}+e^{-i\theta})-2s} e^{in\theta} d\theta$$

$$= \frac{1}{2\pi i} \int_{s^1} \frac{\exp[i\mu(s-rz)]}{z^2-2(s/r)z+1} z^n dz, \qquad (2.7)$$

$$K_n^-(r,s,\mu) = \frac{1}{2\pi} \int_0^{2\pi} \frac{e^{-i\mu(s-re^{-i\theta})}}{r(e^{i\theta}+e^{-i\theta})-2s} e^{in\theta} d\theta$$

$$= \frac{1}{2\pi} \int_0^{2\pi} \frac{e^{-i\mu(s-re^{i\theta})}}{r(e^{i\theta}+e^{-i\theta})-2s} e^{-in\theta} d\theta \qquad (2.8)$$

$$= \frac{1}{2\pi i} \int_{s^1} \frac{\exp[-i\mu(s-rz)]}{z^2-2(s/r)z+1} z^{-n} dz.$$

With such notations, equations (2.5) and (2.6) become

$$f_n(r) = \frac{1}{\pi} \int_{-\infty}^{\infty} K_n^+(r,s,\mu) p_n'(s,\mu) ds$$

$$= \frac{1}{\pi} \int_{-\infty}^{\infty} K_n^-(r,s,\mu) p_n'(s,\mu) ds. \qquad (2.9)$$

For $r > 0$ and $s \in R^1$ we define two functions

$$a_{\pm}(s,r) = \frac{s \pm \sqrt{s^2-r^2}}{r}, \qquad (2.10)$$

$$b_{\pm}(s,r) = \frac{s \pm \text{sgn}(s)\sqrt{s^2-r^2}}{r}. \qquad (2.11)$$

Notice that

$$a_{\pm}^{-n}(s,r) = a_{\mp}^n(s,r) \text{ and } b_{\pm}^{-n}(s,r) = b_{\mp}^n(s,r) \qquad (2.12)$$

$$z^2-2(s/r)z+1 = [z-a_+(s,r)][z-a_-(s,r)] \qquad (2.13)$$

$$= [z-b_+(s,r)][z-b_-(s,r)]$$

To derive closed-form formulas for (2.9) we need evaluate $K_n^+(r,s,\mu)$ and $K_n^-(r,s,\mu)$. For this purpose, we rewrite $K_n^+(r,s,\mu)$ to two different expressions as

$$K_n^+(r,s,\mu) = \frac{1}{2\pi i} \int_{S^1} \frac{\exp[i\mu(s-rz)]r}{2\sqrt{s^2-r^2}} \left(\frac{1}{z-a_+(s,r)} - \frac{1}{z-a_-(s,r)}\right) z^n dz$$

$$= \frac{1}{2\pi i} \int_{S^1} \frac{\exp[i\mu(s-rz)]}{z-b_+(s,r)} \frac{1}{z-b_-(s,r)} z^n dz \qquad (2.14)$$

If $|s| > r$, the integrand of (2.14) has only one pole $b_-(s,r)$ inside the unit disk. If $|s| < r$, the integral of (2.14) contains two Hilbert transforms on the unit circle with singular points of $a_+(s,r)$ and $a_-(s,r)$. It is straightforward to verify the following identities

$$\exp[i\mu(s-ra_\pm(s,r)] = \exp[\mp i\mu\sqrt{s^2-r^2}], \qquad (2.15)$$

$$\exp[-i\mu(s-ra_\pm(s,r)] = \exp[\pm i\mu\sqrt{s^2-r^2}], \qquad (2.16)$$

$$\exp[\pm i\mu(s-rb_-(s,r)] = \exp[\pm i\mu\,\mathrm{sgn}(s)\sqrt{s^2-r^2}] \qquad (2.17)$$

Applying the residual theorem for $|s| > r$ and the Plemelj formulas (A-6) for $|s| < r$, we obtain an explicit expression for $K_n^+(r,s,\mu)$ when $n \geq 0$

$$K_n^+(r,s,\mu) = \frac{1}{2r} \begin{cases} \dfrac{r}{2\sqrt{s^2-r^2}}[e^{-i\mu\sqrt{s^2-r^2}}a_+^n(s,r) - e^{i\mu\sqrt{s^2-r^2}}a_-^n(s,r)] & |s| < r \\[2ex] -\dfrac{\mathrm{sgn}(s)r}{\sqrt{s^2-r^2}}\exp[i\mu\,\mathrm{sgn}(s)\sqrt{s^2-r^2}]b_-^n(s,r) & |s| > r \end{cases}$$

$$= \frac{1}{2r} \begin{cases} \widetilde{U}_{n-1}(s,r,-\mu) & |s| < r \\[1ex] \widetilde{U}_{n-1}(s,r,-\mu) - \dfrac{\mathrm{sgn}(s)r}{\sqrt{s^2-r^2}}\widetilde{T}_n(s,r,-\mu) & |s| > r. \end{cases} \qquad (2.18)$$

We mention that for $n < 0$, $K_n^+(r,s,\mu)$ has an extra pole at $z = 0$, (A-6) is no longer applicable to evaluate $K_n^+(r,s,\mu)$. Our method to obtain $f_n(r)$ for $n < 0$ is to use $K_n^-(r,s,\mu)$. We rewrite $K_n^-(r,s,\mu)$ to

$$K_n^-(r,s,\mu) = \frac{1}{2\pi i} \int_{S^1} \frac{\exp[-i\mu(s-rz)]r}{2\sqrt{s^2-r^2}} \left(\frac{1}{z-a_+(s,r)} - \frac{1}{z-a_-(s,r)}\right) z^{-n} dz$$

$$= \frac{1}{2\pi i} \int_{S^1} \frac{\exp[-i\mu(s-rz)]}{z-b_+(s,r)} \frac{1}{z-b_-(s,r)} z^{-n} dz \qquad (2.19)$$

For $n < 0$, applying the residue theorem and the Plemelj formulas (A-6) to (2.19), using (2.12), (2.16) and (2.17), we obtain an explicit expression

$$K_n^-(r,s,\mu) = \frac{1}{2r} \begin{cases} \dfrac{r}{2\sqrt{s^2-r^2}}[e^{i\mu\sqrt{s^2-r^2}}a_+^{-n}(s,r) - e^{-i\mu\sqrt{s^2-r^2}}a_-^{-n}(s,r)] & |s| < r \\[2ex] -\dfrac{\mathrm{sgn}(s)r}{\sqrt{s^2-r^2}}\exp[-i\mu\,\mathrm{sgn}(s)\sqrt{s^2-r^2}]b_-^{-n}(s,r) & |s| > r \end{cases}$$

$$= \frac{1}{2r} \begin{cases} \dfrac{r}{2\sqrt{s^2-r^2}}[e^{i\mu\sqrt{s^2-r^2}}a_-^n(s,r) - e^{-i\mu\sqrt{s^2-r^2}}a_+^n(s,r)] & |s| < r \\[2ex] -\dfrac{\mathrm{sgn}(s)r}{\sqrt{s^2-r^2}}\exp[-i\mu\,\mathrm{sgn}(s)\sqrt{s^2-r^2}]b_+^n(s,r) & |s| > r \end{cases} \qquad (2.20)$$

$$= \frac{1}{2r} \begin{cases} -\widetilde{U}_{n-1}(s,r,-\mu) & |s| < r \\[1ex] -\widetilde{U}_{n-1}(s,r,-\mu) - \dfrac{\mathrm{sgn}(s)r}{\sqrt{s^2-r^2}}\widetilde{T}_n(s,r,-\mu) & |s| > r \end{cases}$$

Combining (2.18) for $n \geq 0$ and (2.20) for $n < 0$, we have a unified expression as follows

$$\hat{K}_n(r,s,\mu) = \frac{1}{2r}\begin{cases} \mathrm{sgn}(n)\widetilde{U}_{n-1}(s,r,-\mu) & |s| < r \\ \mathrm{sgn}(n)\widetilde{U}_{n-1}(s,r,-\mu) - \dfrac{\mathrm{sgn}(s)r}{\sqrt{s^2 - r^2}}\widetilde{T}_n(s,r,-\mu) & |s| > r. \end{cases} \quad (2.21)$$

Using (2.9), (2.1) can be symbolically written as

$$f_n(r) = \frac{1}{\pi}\int_{-\infty}^{\infty}\hat{K}_n(r,s,\mu)\,p_n'(s,\mu)ds. \quad (2.22)$$

This completes the proof of (2.1). $\qquad\square$

**Remark 1**. *As shown in the derivation of (2.18) and (2.20), functions $K_n^+(r,s,\mu)$ and $K_n^-(r,s,\mu)$ are the Hilbert transforms on the unit circle and can be evaluated by Plemelj formula (A-6). Notice that $K_n^+(r,s,\mu)$ and $K_n^-(r,s,\mu)$ are the same as $I_1(r,s)$ and $I_2(r,s)$ defined in [19], respectively. In [19], $I_1(r,s)$ and $I_2(r,s)$ are evaluated by the generator of Chebyshev polynomials*

$$\frac{1}{z^2 - 2(s/r)z + 1} = \sum_{n\geq 0}U_n(\frac{s}{r})z^n. \quad (2.23)$$

*Series expansion (2.23) converges if $|z| < 1$ and $|s| < r$. For $z \in S^1$, (2.7) and (2.8) are understood as principal integrals and it does not seem to be obvious that (2.23) can be used to calculate the Hilbert transform on the unit circle for singular points $\dfrac{s \pm i\sqrt{r^2 - s^2}}{r} \in S^1$. In this paper formula (4.6) in [19] will be proved by using range condition (1.12) and (A-5) instead of (2.23).*

**Theorem 2**. *Under assumption* (**A1**), *let $p_n'(s,\mu)$ be the derivative of $p_n(s,\mu)$ with respect to $s$, then $f_n(r)$ can be reconstructed by the following formulas*

$$f_n(r) = \frac{1}{2\pi r}\int_{-r}^{r}e^{-i\,\mathrm{sgn}(n)\mu s}\sum_{k=0}^{|n|-1}\frac{(\mathrm{sgn}(n)i\mu r)^k}{k!}\widetilde{U}_{|n|-k-1}(s,r,0)\,p_n'(s,\mu)ds$$
$$-\frac{1}{2\pi r}\int_{|s|>r}\frac{\mathrm{sgn}(s)r}{\sqrt{s^2 - r^2}}\widetilde{T}_n(s,r,-\mu)\,p_n'(s,\mu)ds, \quad (2.24)$$

$$f_n(r) = -\frac{1}{2\pi r}\int_{|s|>r}e^{-i\,\mathrm{sgn}(n)\mu s}\sum_{k=0}^{|n|-1}\frac{(\mathrm{sgn}(n)i\mu r)^k}{k!}\widetilde{U}_{|n|-k-1}(s,r,0)\,p_n'(s,\mu)ds$$
$$-\frac{1}{2\pi r}\int_{|s|>r}\frac{\mathrm{sgn}(s)r}{\sqrt{s^2 - r^2}}\widetilde{T}_n(s,r,-\mu)\,p_n'(s,\mu)ds. \quad (2.25)$$

*Moreover $p_n'(s,\mu)$ satisfies the following range condition*

$$\int_{-\infty}^{\infty}\widetilde{U}_{n-1}(s,r,-\mu)\,p_n'(s,\mu)ds = \int_{-r}^{r}e^{-i\,\mathrm{sgn}(n)\mu s}\sum_{k=0}^{|n|-1}\frac{(i\,\mathrm{sgn}(n)\mu r)^k}{k!}\widetilde{U}_{|n|-k-1}(s,r,0)\,p_n'(s,\mu)ds. \quad (2.26)$$

**Proof**. Formulas (2.24) was previously derived in [19] with help of (2.23). In this paper we use (A-5) to derive (2.24). From (2.9) we have

$$f_n(r) = \frac{1}{2\pi}\int_{-\infty}^{\infty}[K_n^+(r,s,\mu) + K_n^-(r,s,\mu)]\,p_n'(s,\mu)ds \quad (2.27)$$

The main task is to obtain an explicit expression of (2.27) through evaluating $K_n^+(r,s,\mu)$ for $n < 0$ and $K_n^-(r,s,\mu)$ for $n \geq 0$.

First we consider $K_n^+(r,s,\mu)$ for $n < 0$. In case of $|s| < r$, we rewrite (2.7) to

$$K_n^+(r,s,\mu) = \frac{1}{2\pi i} \int_{S^1} \frac{\exp(i\mu s)}{z^2 - 2(s/r)z + 1} \left[\sum_{k=0}^{|n|-1} \frac{1}{k!}(-i\mu rz)^k + \sum_{k=|n|}^{\infty} \frac{1}{k!}(-i\mu rz)^k\right] z^n dz$$

$$= \frac{1}{2\pi i} \int_{S^1} \frac{\exp(i\mu s)}{z^2 - 2(s/r)z + 1} \left[\sum_{k=0}^{|n|-1} \frac{(-i\mu r)^k}{k!} z^{k+n} + \sum_{k=|n|}^{\infty} \frac{(-i\mu r)^k}{k!} z^{k+n}\right] dz \qquad (2.28)$$

Notice that $\sum_{k=0}^{|n|-1} \frac{(-i\mu r)^k}{k!} z^{k+n} + \sum_{k=|n|}^{\infty} \frac{(-i\mu r)^k}{k!} z^{k+n}$ is the series expansion of $\exp(-i\mu r)$, by (A-5) we obtain

$$K_n^+(r,s,\mu)$$

$$= \frac{\exp(i\mu s)r}{4\pi i r\sqrt{s^2-r^2}} \int_{S^1} \left(\frac{1}{z-a_+(s,r)} - \frac{1}{z-a_-(s,r)}\right)\left[\sum_{k=0}^{|n|-1} \frac{(-i\mu r)^k}{k!} z^{k+n} + \sum_{k=|n|}^{\infty} \frac{(-i\mu r)^k}{k!} z^{k+n}\right] dz$$

$$= \frac{\exp(i\mu s)r}{4r\sqrt{s^2-r^2}} \left\{\sum_{k=|n|}^{\infty} \frac{(-i\mu r)^k}{k!}[a_+^{k+n}(s,r) - a_-^{k+n}(s,r)] - \sum_{k=0}^{|n|-1} \frac{(-i\mu r)^k}{k!}[a_+^{k+n}(s,r) - a_-^{k+n}(s,r)]\right\}$$

$$= \frac{\exp(i\mu s)r}{4r\sqrt{s^2-r^2}} \left\{\sum_{k=0}^{\infty} \frac{(-i\mu r)^k}{k!}[a_+^{k+n}(s,r) - a_-^{k+n}(s,r)] - 2\sum_{k=0}^{|n|-1} \frac{(-i\mu r)^k}{k!}[a_+^{k+n}(s,r) - a_-^{k+n}(s,r)]\right\} \qquad (2.29)$$

$$= \frac{\exp(i\mu s)r}{4r\sqrt{s^2-r^2}} \left\{[e^{-i\mu r a_+(s,r)} a_+^n(s,r) - e^{-i\mu r a_-(s,r)} a_-^n(s,r)] - \frac{4\sqrt{s^2-r^2}}{r} \sum_{k=0}^{|n|-1} \frac{(-i\mu r)^k}{k!} \widetilde{U}_{k+n-1}(s,r,0)\right\}$$

$$= \frac{\exp(i\mu s)r}{4r\sqrt{s^2-r^2}} \left\{e^{-i\mu s}[e^{-i\mu\sqrt{s^2-r^2}} a_+^n(s,r) - e^{i\mu\sqrt{s^2-r^2}} a_-^n(s,r)] + \frac{4\sqrt{s^2-r^2}}{r} \sum_{k=0}^{|n|-1} \frac{(-i\mu r)^k}{k!} \widetilde{U}_{|n|-k-1}(s,r,0)\right\}$$

$$= \frac{1}{2r}[\widetilde{U}_{n-1}(s,r,-\mu) + 2e^{i\mu s}\sum_{k=0}^{|n|-1} \frac{(-i\mu r)^k}{k!} \widetilde{U}_{|n|-k-1}(s,r,0)].$$

In case of $|s|>r$, $0$ and $b_-(s,r)$ are the poles inside the unit circle. If $n<-1$, by the residue theorem we have

$$K_n^+(r,s,\mu) = \frac{1}{2\pi i} \int_{S^1} \frac{\exp[i\mu(s-rz)]}{z-b_+(s,r)} \frac{1}{z-b_-(s,r)} z^n dz$$

$$= \frac{1}{r} \frac{\exp[i\mu(s-rb_-(s,r))]}{b_-(s,r) - b_+(s,r)} b_-^n(s,r) \qquad (2.30)$$

$$= -\frac{\operatorname{sgn}(s)}{2\sqrt{s^2-r^2}} \exp[i\mu\operatorname{sgn}(s)\sqrt{s^2-r^2}] b_-^n(s,r)$$

If $n=-1$, we have an extra term $\exp(-i\mu s)$ because of the pole at $0$

$$K_{-1}^+(r,s,\mu) = -\frac{\operatorname{sgn}(s)}{2\sqrt{s^2-r^2}} \exp[i\mu\operatorname{sgn}(s)\sqrt{s^2-r^2}] b_-^n(s,r) + \exp(i\mu s) . \qquad (2.31)$$

By range condition (1.12) with $m=0$, $\exp(i\mu s)$ has no contribution to (2.9). In summary $K_n^+(r,s,\mu)$ of (2.9) for $n<0$ can be expressed as

$$K_n^+(r,s,\mu) = \frac{1}{2r} \begin{cases} \widetilde{U}_{n-1}(s,r,-\mu) + 2e^{-i\operatorname{sgn}(n)\mu s}\sum_{k=0}^{|n|-1} \frac{(\operatorname{sgn}(n)i\mu r)^k}{k!} \widetilde{U}_{|n|-k-1}(s,r,0) & |s|<r \\ \widetilde{U}_{n-1}(s,r,-\mu) - \frac{\operatorname{sgn}(s)r}{\sqrt{s^2-r^2}} \widetilde{T}_n(s,r,-\mu) & |s|>r \end{cases} \qquad (2.32)$$

Next we consider $K_n^-(r,s,\mu)$ for $n\geq 0$. If $|s|<r$ we rewrite (2.8) to

$$K_n^-(r,s,\mu) = \frac{1}{2\pi i} \int_{S^1} \frac{\exp(-i\mu s)}{z^2 - 2(s/r)z + 1} \left[ \sum_{k=0}^{n-1} \frac{1}{k!}(i\mu rz)^k + \sum_{k=n}^{\infty} \frac{1}{k!}(i\mu rz)^k \right] z^{-n} dz$$

$$= \frac{1}{2\pi i} \int_{S^1} \frac{\exp(-i\mu s)}{z^2 - 2(s/r)z + 1} \left[ \sum_{k=0}^{n-1} \frac{(i\mu r)^k}{k!} z^{k-n} + \sum_{k=n}^{\infty} \frac{(i\mu r)^k}{k!} z^{k-n} \right] dz \qquad (2.33)$$

Notice that $\left[ \sum_{k=0}^{n-1} \frac{(i\mu r)^k}{k!} z^{k-n} + \sum_{k=n}^{\infty} \frac{(i\mu r)^k}{k!} z^{k-n} \right]$ is in the format of series expansion, by (A-5) we obtain

$$K_n^-(r,s,\mu)$$

$$= \frac{\exp(-i\mu s)r}{4\pi i r\sqrt{s^2-r^2}} \int_{S^1} \left( \frac{1}{z-a_+(s,r)} - \frac{1}{z-a_-(s,r)} \right) \left[ \sum_{k=0}^{n-1} \frac{(i\mu r)^k}{k!} z^{k-n} + \sum_{k=n}^{\infty} \frac{(i\mu r)^k}{k!} z^{k-n} \right] dz$$

$$= \frac{\exp(-i\mu s)r}{4r\sqrt{s^2-r^2}} \left\{ \sum_{k=n}^{\infty} \frac{(i\mu r)^k}{k!} [a_+^{k-n}(s,r) - a_-^{k-n}(s,r)] - \sum_{k=0}^{n-1} \frac{(i\mu r)^k}{k!} [a_+^{k-n}(s,r) - a_-^{k-n}(s,r)] \right\}$$

$$= \frac{\exp(-i\mu s)r}{4r\sqrt{s^2-r^2}} \left\{ e^{i\mu r a_+(s,r)} a_+^{-n}(s,r) - e^{i\mu r a_-(s,r)} a_-^{-n}(s,r) - \frac{4\sqrt{s^2-r^2}}{r} \sum_{k=0}^{n-1} \frac{(i\mu r)^k}{k!} \widetilde{U}_{k-n-1}(s,r,0) \right\} \qquad (2.34)$$

$$= \frac{\exp(-i\mu s)r}{4r\sqrt{s^2-r^2}} \left\{ e^{i\mu s} [e^{i\mu\sqrt{s^2-r^2}} a_-^n(s,r) - e^{-i\mu\sqrt{s^2-r^2}} a_+^n(s,r)] + \frac{4\sqrt{s^2-r^2}}{r} \sum_{k=0}^{|n|-1} \frac{(i\mu r)^k}{k!} \widetilde{U}_{|n|-k-1}(s,r,0) \right\}$$

$$= \frac{1}{2r} [-\widetilde{U}_{n-1}(s,r,-\mu) + 2e^{-i\mu s} \sum_{k=0}^{|n|-1} \frac{(i\mu r)^k}{k!} \widetilde{U}_{|n|-k-1}(s,r,0)].$$

If $|s| > r$, $0$ and $b_-(s,r)$ are two poles inside the unit circle. If $n \neq 1$, by the residue theorem we have

$$K_n^-(r,s,\mu) = \frac{1}{2\pi i} \int_{S^1} \frac{\exp[-i\mu(s-rz)]}{z^2 - 2(s/r)z + 1} z^{-n} dz$$

$$= \frac{1}{2\pi i} \int_{S^1} \frac{\exp[-i\mu(s-rz)]}{z-b_+(s,r)} \frac{1}{z-b_-(s,r)} \frac{1}{z^n} dz$$

$$= \frac{1}{r} \int_{S^1} \frac{\exp[-i\mu(s-rb_-(s,r))]}{b_-(s,r) - b_+(s,r)} b_-^{-n}(s,r) dz \qquad (2.35)$$

$$= -\frac{\text{sgn}(s)}{2\sqrt{s^2-r^2}} \exp[-i\mu \text{sgn}(s)\sqrt{s^2-r^2}] b_-^{-n}(s,r)$$

For the case of $n=1$, we have an extra term $\exp(-i\mu s)$ because of the pole at $0$

$$K_1^-(r,s,\mu) = -\frac{\text{sgn}(s)}{2\sqrt{s^2-r^2}} \exp[-i\mu \text{sgn}(s)\sqrt{s^2-r^2}] b_-^{-n}(s,r) + \exp(-i\mu s). \qquad (2.36)$$

By range condition (1.12) with $m=0$, $\exp(-i\mu s)$ has no contribution to (2.9). In summary $K_n^-(r,s,\mu)$ of (2.9) for $n \geq 0$ can be defined as

$$K_n^-(r,s,\mu) = \frac{1}{2r} \begin{cases} -\widetilde{U}_{n-1}(s,r,-\mu) + 2e^{-i\,\text{sgn}(n)\mu s} \sum_{k=0}^{|n|-1} \frac{(i\mu r)^k}{k!} \widetilde{U}_{|n|-k-1}(s,r,0) & |s| < r \\ -\widetilde{U}_{n-1}(s,r,-\mu) - \frac{\text{sgn}(s)r}{\sqrt{s^2-r^2}} \widetilde{T}_n(s,r,-\mu) & |s| > r \end{cases} \qquad (2.37)$$

Combining (2.18), (2.20), (2.32) and (2.37), for all integers $n$, we obtain

$$K_n^+(r,s,\mu) + K_n^-(r,s,\mu) = \frac{1}{r} \begin{cases} e^{-i\,\text{sgn}(n)\mu s} \sum_{k=0}^{|n|-1} \frac{(\text{sgn}(n)i\mu r)^k}{k!} U_{|n|-k-1}(s,r,0) & |s| < r \\ -\frac{\text{sgn}(s)r}{\sqrt{s^2-r^2}} \widetilde{T}_n(s,r,-\mu) & |s| > r \end{cases} \qquad (2.38)$$

$$K_n^+(r,s,\mu) - K_n^-(r,s,\mu) = \frac{1}{r}\begin{cases} \widetilde{U}_{n-1}(s,r,-\mu) - e^{-i\operatorname{sgn}(n)\mu s}\sum_{k=0}^{|n|-1}\dfrac{(\operatorname{sgn}(n)i\mu r)^k}{k!}U_{|n|-k-1}(s,r,0) & |s|<r \\ \widetilde{U}_{n-1}(s,r,-\mu) & |s|>r \end{cases} \quad (2.39)$$

The combination of (2.27) and (2.38) leads to (2.24). Using range condition (1.12), (2.24) can be rewritten as (2.25). Identity (2.26) is derived by combining (2.9) and (2.39). This completes the proof. □

**Remark 2**. *Recently (2.24) was extended to radius-dependent attenuation in [20] but some extra terms are needed to reflect the differentials of backprojection weights. We have not been able to derive simpler formulas like (2.24) by using Plemelj formula.*

## III.  Exterior Inversion Formula of (1.9)

For pure imaginary $\mu$, in [19] Puro obtained the following exterior formula

$$f_n(r) = -\frac{1}{\pi}\int_r^\infty \frac{Tc_n(s/r,\beta r)}{\sqrt{s^2-r^2}}\frac{\partial}{\partial s}g_n(s,\beta)ds. \qquad (3.1)$$

The derivation of (3.1) is through (5.2) of [19]. Following the same idea used in [19], we prove that (3.1) holds for complex constant $\mu$.

**Lemma 1** (Orthogonality of exponential Chebyshev functions). *For $0<s<t$ and $\mu\in C$, we have*

$$\int_s^t \frac{\widetilde{T}_n(t,r,\mu)}{\sqrt{t^2-r^2}}\frac{\widetilde{T}_n(s,r,\mu)}{\sqrt{r^2-s^2}}r\,dr = \frac{\pi}{2} \qquad (3.2)$$

**Proof**. It is easy to verify the following equation

$$4\widetilde{T}_n(t,r,\mu)\widetilde{T}_n(s,r,\mu)$$
$$= e^{i\mu(\sqrt{t^2-r^2}+\sqrt{s^2-r^2})}\left(\frac{t+\sqrt{t^2-r^2}}{s-\sqrt{s^2-r^2}}\right)^n + e^{-i\mu(\sqrt{t^2-r^2}+\sqrt{s^2-r^2})}\left(\frac{t-\sqrt{t^2-r^2}}{s+\sqrt{s^2-r^2}}\right)^n \qquad (3.3)$$
$$+ e^{i\mu(\sqrt{t^2-r^2}-\sqrt{s^2-r^2})}\left(\frac{t+\sqrt{t^2-r^2}}{s+\sqrt{s^2-r^2}}\right)^n + e^{-i\mu(\sqrt{t^2-r^2}-\sqrt{s^2-r^2})}\left(\frac{t-\sqrt{t^2-r^2}}{s-\sqrt{s^2-r^2}}\right)^n.$$

In $C\setminus[-t,t]$, we define a complex function

$$\Phi_n(z,s,t,\mu) = \frac{z}{\sqrt{s^2-z^2}\sqrt{t^2-z^2}}$$
$$\times[e^{i\mu(\sqrt{t^2-z^2}-\sqrt{s^2-z^2})}\left(\frac{t+\sqrt{t^2-z^2}}{s+\sqrt{s^2-z^2}}\right)^n + e^{-i\mu(\sqrt{t^2-z^2}-\sqrt{s^2-z^2})}\left(\frac{t-\sqrt{t^2-z^2}}{s-\sqrt{s^2-z^2}}\right)^n]. \qquad (3.4)$$

In [19] taking the real part of the complex function was used. The definition of (3.4) allows us to avoid taking the real part of $\Phi_n(z,s,t,\mu)$ in calculating the path integrals. We chose the branch cuts of $\sqrt{t^2-z^2}$ and $\sqrt{s^2-z^2}$ such that $\Phi_n(z,s,t,\mu)$ is analytical in $C\setminus([-s,-t]\cup[s,t])$ and

$$\lim_{|z|\to\infty}\frac{\sqrt{t^2-z^2}}{z} = \lim_{|z|\to\infty}\frac{\sqrt{s^2-z^2}}{z} = i, \qquad (3.5)$$

$$\lim_{\varepsilon>0,\varepsilon\to0}\sqrt{t^2-(r\pm i\varepsilon)^2} = \pm\sqrt{t^2-r^2}, \quad s\triangleleft r|<t, \qquad (3.6)$$

$$\lim_{\varepsilon>0,\varepsilon\to0}\sqrt{s^2-(r\pm i\varepsilon)^2} = \operatorname{sign}(r)\sqrt{s^2-r^2}, \quad s\triangleleft r|<t. \qquad (3.7)$$

The cuts are described in Fig. 1:

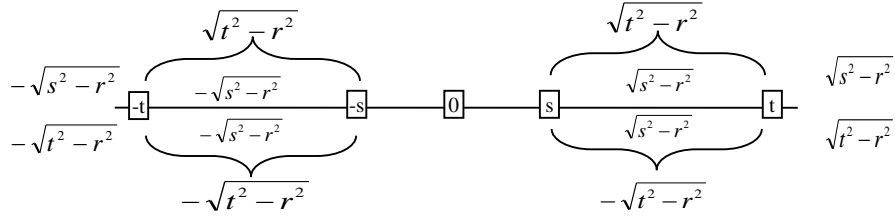

Fig. 1 Illustration of cuts for $\sqrt{t^2 - z^2}$ and $\sqrt{s^2 - z^2}$.

Taking the path integral on the large unit circle along the clockwise orientation, we obtain

$$\lim_{|z| \to \infty} \Phi_n(z, s, t, \mu) z = -2, \quad \lim_{d \to \infty} \oint_{|z| = d} \Phi_n(z, s, t, \mu) dz = 4\pi i. \tag{3.8}$$

It follows that

$$\lim_{\substack{\varepsilon > 0 \\ \varepsilon \to 0}} \left\{ \int_{-t}^{-s} [\Phi_n(r + i\varepsilon, s, t, \mu) - \Phi_n(r - i\varepsilon, s, t, \mu)] dr + \int_{s}^{t} [\Phi_n(r + i\varepsilon, s, t, \mu) - \Phi_n(r - i\varepsilon, s, t, \mu)] dr \right\} \tag{3.9}$$

$$= 4\pi i.$$

For $r \in (s, t)$, we have

$$\lim_{\substack{\varepsilon > 0 \\ \varepsilon \to 0}} [\Phi_n(r + i\varepsilon, s, t, \mu) - \Phi_n(r - i\varepsilon, s, t, \mu)]$$

$$= \frac{r}{\sqrt{t^2 - r^2} \sqrt{s^2 - r^2}} [e^{i\mu(\sqrt{t^2 - r^2} - \sqrt{s^2 - r^2})} \left( \frac{t + \sqrt{t^2 - r^2}}{s + \sqrt{s^2 - r^2}} \right)^n + e^{-i\mu(\sqrt{t^2 - r^2} - \sqrt{s^2 - r^2})} \left( \frac{t - \sqrt{t^2 - r^2}}{s - \sqrt{s^2 - r^2}} \right)^n] \tag{3.10}$$

$$+ \frac{r}{\sqrt{t^2 - r^2} \sqrt{s^2 - r^2}} [e^{-i\mu(\sqrt{t^2 - r^2} + \sqrt{s^2 - r^2})} \left( \frac{t - \sqrt{t^2 - r^2}}{s + \sqrt{s^2 - r^2}} \right)^n + e^{i\mu(\sqrt{t^2 - r^2} + \sqrt{s^2 - r^2})} \left( \frac{t + \sqrt{t^2 - r^2}}{s - \sqrt{s^2 - r^2}} \right)^n].$$

For $r \in (-t, -s)$, we have

$$\lim_{\substack{\varepsilon > 0 \\ \varepsilon \to 0}} [\Phi_n(r + i\varepsilon, s, t, \mu) - \Phi_n(r - i\varepsilon, s, t, \mu)]$$

$$= \frac{-r}{\sqrt{t^2 - r^2} \sqrt{s^2 - r^2}} [e^{i\mu(\sqrt{t^2 - r^2} + \sqrt{s^2 - r^2})} \left( \frac{t + \sqrt{t^2 - r^2}}{s - \sqrt{s^2 - r^2}} \right)^n + e^{-i\mu(\sqrt{t^2 - r^2} + \sqrt{s^2 - r^2})} \left( \frac{t - \sqrt{t^2 - r^2}}{s + \sqrt{s^2 - r^2}} \right)^n] \tag{3.11}$$

$$- \frac{r}{\sqrt{t^2 - r^2} \sqrt{s^2 - r^2}} [e^{-i\mu(\sqrt{t^2 - r^2} - \sqrt{s^2 - r^2})} \left( \frac{t - \sqrt{t^2 - r^2}}{s - \sqrt{s^2 - r^2}} \right)^n + e^{i\mu(\sqrt{t^2 - r^2} - \sqrt{s^2 - r^2})} \left( \frac{t + \sqrt{t^2 - r^2}}{s + \sqrt{s^2 - r^2}} \right)^n].$$

Combining equations from (3.9) to (3.11), we obtain

$$\int_{s}^{t} \frac{\tilde{T}_n(t, r, \mu)}{\sqrt{t^2 - r^2}} \frac{\tilde{T}_n(s, r, \mu)}{\sqrt{r^2 - s^2}} r dr$$

$$= \frac{1}{i} \int_{s}^{t} \frac{\tilde{T}_n(t, r, \mu)}{\sqrt{t^2 - r^2}} \frac{\tilde{T}_n(s, r, \mu)}{\sqrt{s^2 - r^2}} r dr$$

$$= \frac{1}{8i} \lim_{\substack{\varepsilon > 0 \\ \varepsilon \to 0}} \left\{ \int_{-t}^{-s} [\Phi_n(r + i\varepsilon, s, t, \mu) - \Phi_n(r - i\varepsilon, s, t, \mu)] dr + \int_{s}^{t} [\Phi_n(r + i\varepsilon, s, t, \mu) - \Phi_n(r - i\varepsilon, s, t, \mu)] dr \right\} \tag{3.12}$$

$$= \frac{\pi}{2}$$

This completes the proof. □

In the case of $\mu \in R^1$, from [28-31], $f(r,\varphi)$ can be reconstructed from $p(s,\theta,\mu)$ in $R^1 \times [0,\pi]$, which is often called the half-scan in SPECT. To our knowledge, no result has been obtained to reconstruct $f(r,\varphi)$ from $p(s,\theta,\mu)$ in $R^1_+ \times [0,2\pi]$ or $R^1_- \times [0,2\pi]$. For $s > 0$, $p_n(s,\mu)$ can be obtained from $p(s,\theta,\mu)$ in $R^1_+ \times [0,2\pi]$, and $p_n(s,-\mu)$ can be calculated from $p(s,\theta,\mu)$ in $R^1_- \times [0,2\pi]$ by (1.10). Here we prove that (3.1) holds for complex $\mu$.

**Theorem 3**. *If* $p(s,\theta,\mu)$ *is a smooth function in* $\bar{I} \times S^1$. *Define* $f_n(r)$ *as*

$$f_n(r) = -\frac{1}{\pi} \int_r^\infty \frac{\tilde{T}_n(t,r,\mu)}{\sqrt{t^2 - r^2}} p'_n(t,\mu) dt , \tag{3.13}$$

*Then* $f_n(r)$ *satisfy* (1.9) *in* $R^1_+ \times [0,2\pi]$.

**Proof**. The proof of (3.13) uses the same arguments in [5, 19] as follows

$$2\int_s^\infty \frac{\tilde{T}_n(s,r,\mu)}{\sqrt{r^2 - s^2}} f_n(r) r\, dr$$

$$= -\frac{2}{\pi} \int_s^\infty \frac{\tilde{T}_n(s,r,\mu)}{\sqrt{r^2 - s^2}} [\int_r^\infty \frac{\tilde{T}_n(t,r,\mu)}{\sqrt{t^2 - r^2}} p'_n(t,\mu) dt] r\, dr \tag{3.14}$$

$$= -\frac{2}{\pi} \int_s^\infty [\int_s^t \frac{\tilde{T}_n(t,r,\mu)}{\sqrt{t^2 - r^2}} \frac{\tilde{T}_n(s,r,\mu)}{\sqrt{r^2 - s^2}} r\, dr] p'_n(t,\mu) dt$$

$$= p_n(s,\mu).$$

Thus the ERT of $f(r,\varphi) = \sum_{-\infty}^\infty f_n(r) e^{in\varphi}$ is $p(s,\theta,\mu)$ on $R^1_+ \times [0,2\pi]$. This completes the proof. □

Puro and Garin extended some inversion formulas of [19] for the imaginary attenuation to the radius-dependent complex attenuation in [20]. Hereafter $\eta(r)$ denotes the radius-dependent complex coefficients. To be consistent with the notations used in this paper, we reformulate several definitions of [20] as follows

$$a(s,r,\eta) = \int_0^{\sqrt{r^2 - s^2}} \eta(\sqrt{s^2 + t^2}) dt = i\sqrt{s^2 - r^2} \int_0^1 \eta(\sqrt{(1-\tau^2)s^2 + \tau^2 r^2}) d\tau , \tag{3.15}$$

$$b(s,\eta) = \int_0^{|s|} \frac{\eta(\rho)\rho}{s\sqrt{1 - (\rho/s)^2}} d\rho = s \int_0^1 \frac{\eta(s\rho)\rho}{\sqrt{1 - \rho^2}} d\rho , \tag{3.16}$$

$$\tilde{T}_n(s,r,\eta) = \frac{1}{2} [e^{a(s,r,\eta)} (\frac{s + \sqrt{s^2 - r^2}}{r})^n + e^{-a(s,r,\eta)} (\frac{s - \sqrt{s^2 - r^2}}{r})^n] , \tag{3.17}$$

$$\tilde{U}_{n-1}(s,r,\eta) = \frac{r}{2\sqrt{s^2 - r^2}} [e^{a(s,r,\eta)} (\frac{s + \sqrt{s^2 - r^2}}{r})^n - e^{-a(s,r,\eta)} (\frac{s - \sqrt{s^2 - r^2}}{r})^n] , \tag{3.18}$$

$$p(s,\theta,\eta) = \int_{-\infty}^\infty f(s\bar{\theta} + t\bar{\theta}^\perp) e^{\text{sgn}(t)a(s,\sqrt{s^2 + t^2},\eta)} dt , \tag{3.19}$$

$$p(s,\theta,\eta) = \sum_{-\infty}^\infty p_n(s,\eta) e^{in\theta} , \tag{3.20}$$

$$p_n(s,\eta) = 2\int_{|s|}^\infty \frac{\tilde{T}_n(s,r,\eta)}{\sqrt{r^2 - s^2}} f_n(r) r\, dr . \tag{3.21}$$

Here we point out that both (3.15) and (3.16) are changed to path integrals so that $a(s,r,\eta)$ and $b(s,\eta)$ can be easily extended to complex functions in $C$ if $\eta(r)$ is extended as a complex function. Next we present our theorem 4 with sketchy proof.

**Theorem 4**. *Assume that $\eta(r)$ is a complex smooth function on $\bar{I}$ and can be extended to an analytical function $\eta(z)$ in $C\backslash[-t,t]$ and continuous on $C$ with the following growth condition*

$$|a(t,z,\eta)-a(s,z,\eta)|\to 0 \ \ when \ |z|\to\infty. \tag{3.22}$$

*Then for $0<s<t$, we have*

$$\int_s^t \frac{\widetilde{T}_n(t,r,\eta)}{\sqrt{t^2-r^2}}\frac{\widetilde{T}_n(s,r,\eta)}{\sqrt{r^2-s^2}}rdr = \frac{\pi}{2}, \tag{3.23}$$

*and $f_n(r)$ satisfies (3.21) in $R_+^1\times[0,2\pi]$, here $f_n(r)$ is defined as*

$$f_n(r) = -\frac{1}{\pi}\int_r^\infty \frac{\widetilde{T}_n(t,r,\eta)}{\sqrt{t^2-r^2}}p_n'(t,\eta)dt. \tag{3.24}$$

**Proof**. The arguments mainly follow the proof of Lemma 1 and Theorem 3. We only provide the key steps. From the assumption, $\eta(r)$ can be understood as the restriction of a continuous function $\eta(z)$ on $\bar{I}$, thus $\eta(z)$ is continuous in the region that contains $[-t,t]$. In $C\backslash[-t,t]$, we modify the complex function of (3.4) as

$$\Psi_n(z,s,t,\eta) = \frac{z}{\sqrt{s^2-z^2}\sqrt{t^2-z^2}}$$
$$\times [e^{a(t,z,\eta)-a(s,z,\eta)}\left(\frac{t+\sqrt{t^2-z^2}}{s+\sqrt{s^2-z^2}}\right)^n + e^{-[a(t,z,\eta)-a(s,z,\eta)]}\left(\frac{t-\sqrt{t^2-z^2}}{s-\sqrt{s^2-z^2}}\right)^n]. \tag{3.25}$$

We choose the same cuts as for $\Psi_n(z,s,t,\eta)$ satisfying conditions from (3.5) to (3.7). The growth condition (3.22) and continuity of $\eta(z)$ on $C$ yield (3.8) and (3.12) for $\Psi_n(z,s,t,\eta)$. By using the same arguments in the proof of Lemma 1 and Theorem 3, we obtain (3.23) and (3.24). This completes the proof. □

According to the calculations in [20], Novikov's inversion formula can be rewritten as

$$f(r,\varphi) = \frac{1}{4\pi^2}\int_0^{2\pi}\frac{\partial}{\partial l}[e^{-a(l,r,\eta)}\int_{-\infty}^\infty \frac{\cos(b(l,\eta)-b(s,\eta))}{l-s}p(\theta,s,\eta)ds]|_{l=\vec{r}\cdot\vec{\theta}}\,d\theta. \tag{3.26}$$

Compared with (2.2, 2.3), (3.26) requires evaluating the differentials of $a(l,r,\eta)$ and $b(s,\eta)$. If $\eta(r)$ is not a constant, (2.2) and (2.3) don't hold due to the differentials of the backprojection weights $e^{-a(l,r,\eta)}$ and convolutional kernel $\cos(b(l,\eta)-b(s,\eta))$, then formula (3.10) of [20] includes extra terms. In [20], the generating series (2.23) was used to calculate the singular integrals in the derivation. After some calculations, we have not been able to obtain simplified inversion formulas like (2.1) and (2.24) for the radius-dependent complex coefficients $\eta(r)$.

# IV. Range Conditions

A review on the range conditions for the attenuated Radon transform can be found in [6] and the references therein. The classic works on the range condition for the Radon transform are from [32, 33]. The range condition includes the evenness condition (1.10) and the moment condition (1.11). We point out that (1.11) is new for $r>0$. The sufficiency of the combination of (1.10) and (1.11) has been proved for the case of $\mu = 0$. However, if $\mu \neq 0$, it has not been proved that the evenness condition (1.10) and moment condition (1.11) are sufficient to uniquely reconstruct $f(r, \varphi)$ from $p(s, \theta, \mu)$. Finch mentioned that it is desirable to have a simple proof to show that Novikov's range condition can yield the existing range conditions for the (exponential) Radon transform. In this paper, for $\mu \in R^1$, we will prove that Novikov's range condition does imply the existing range conditions.

Recall that Novikov's range condition is expressed in an integral identity by

$$\int_0^{2\pi} e^{-\mu \vec{r} \cdot \vec{\theta}^\perp} \left[ \int_{-\infty}^{\infty} \frac{\cosh(i\mu(\vec{r} \cdot \vec{\theta} - s))}{\vec{r} \cdot \vec{\theta} - s} p(\theta, s, \mu) ds \right] d\theta = 0 . \tag{4.1}$$

A direct proof of (4.1) can be found in [4] and the sufficiency of (4.1) was proved in [12] in the context of attenuated Radon transform. Subtracting (2.3) from (2.2), we obtain

$$\int_0^{2\pi} e^{-\mu \vec{r} \cdot \vec{\theta}^\perp} \left[ \int_{-\infty}^{\infty} \frac{\sinh(i\mu(\vec{r} \cdot \vec{\theta} - s))}{\vec{r} \cdot \vec{\theta} - s} p'(\theta, s, \mu) ds \right] d\theta = 0 . \tag{4.2}$$

From the proof of Theorem 2, (2.26) is the circular harmonic expression of (4.2). So far, we have three range conditions of (1.11), (4.1) and (4.2) derived through different methods. It may be interesting to investigate the relations among these range conditions. In the circular harmonic expression, by (2.38), Novikov's condition (4.1) becomes

$$\int_{|s|>r} \text{sgn}(s) \frac{\tilde{T}_n(s, r, -\mu)r}{\sqrt{s^2 - r^2}} p_n(s, \mu) ds = \int_{-r}^{r} e^{-i\,\text{sgn}(n)\mu s} \sum_{k=0}^{|n|-1} \frac{(i\,\text{sgn}(n)\mu r)^k}{k!} U_{|n|-k-1}(s, r, 0) p_n(s, \mu) ds . \tag{4.3}$$

Under condition (**A1**), if $r > 1$, the left hand side of (4.3) becomes zero, then we have

$$\int_{-1}^{1} \left[ e^{-i\,\text{sgn}(n)\mu s} \sum_{k=0}^{|n|-1} \frac{(i\,\text{sgn}(n)\mu r)^k}{k!} U_{|n|-k-1}(s, r, 0) \right] p_n(s, \mu) ds = 0 . \tag{4.4}$$

We rewrite (4.4) in the series of $r^q$, $|n| - 1 \geq q \geq 1 - |n|$, as follows

$$\sum_{q=1-|n|}^{|n|-1} c_q r^q = 0 , \ r > 1 . \tag{4.5}$$

It follows that $c_q = 0$. Comparing the coefficients in (4.4), if $\mu \neq 0$, we obtain

$$\int_{-1}^{1} s^m e^{-i\,\text{sgn}(n)\mu s} p_n(s, \mu) ds = 0 , \ 0 \leq m < |n| . \tag{4.6}$$

This implies that under condition (**A1**), (4.1) leads to the moment condition (1.11) for $r = 0$.

Let $\tilde{p}(\omega, \theta, \mu)$ be the Fourier transform of $p(s, \theta, \mu)$ with respect to $s$

$$\tilde{p}(\omega, \theta, \mu) = \int_{-\infty}^{\infty} p(s, \theta, \mu) e^{-is\omega} ds . \tag{4.7}$$

Expanding $\tilde{p}(\omega, \theta, \mu)$ in its circular harmonics, we have

$$\tilde{p}(\omega, \theta, \mu) = \sum_{-\infty}^{\infty} \tilde{p}_n(\omega, \mu) e^{in\theta} . \tag{4.8}$$

From [7, 8] we have the following evenness identity

$$(\mu + \omega)^n \tilde{p}_n(\omega, \mu) = (\mu - \omega)^n \tilde{p}_n(-\omega, \mu) , \ |\omega| > |\mu| . \tag{4.9}$$

The sufficiency of (4.9) was proved in [10] for $\mu \in R^1$. Next we prove that for $\mu \in R^1$, (4.1) does yield the evenness condition (4.9).

**Theorem 5** *Assume that $\mu \in R^1$, $p(s, \theta, \mu)$ meets (4.1) and $p(s, \theta, \mu)$ is continuously differentiable with support in $\bar{I} \times S^1$, then $\tilde{p}_n(\omega, \mu)$ satisfies (4.9).*

**Proof**. In the distribution sense, the Fourier transforms of $e^{\pm i\mu s}$ and $\frac{1}{\pi s}$ are $\sqrt{2\pi}\delta(\omega \mp \mu)$ and $-\frac{i\,\mathrm{sgn}(\omega)}{\sqrt{2\pi}}$, respectively. The convolution of $\delta(\omega \mp \mu)$ and $\mathrm{sgn}(\omega)$ is $\mathrm{sgn}(\omega \mp \mu)$. Using the polar coordinate expressions of $\vec{r}$, $\vec{r}\cdot\vec{\theta} = r\cos(\theta-\varphi)$ and $-\vec{r}\cdot\vec{\theta}^{\perp} = r\sin(\theta-\varphi)$, we obtain

$$
\int_{-\infty}^{\infty} \frac{\cosh(i\mu(\vec{r}\cdot\vec{\theta}-s))}{\vec{r}\cdot\vec{\theta}-s}\, p(\theta,s,\mu)\,ds
$$
$$
= -\frac{i}{2}\int_{-\infty}^{\infty} (\mathrm{sgn}(\omega+\mu)+\mathrm{sgn}(\omega-\mu))\tilde{p}(\theta,\omega,\mu)e^{i\omega r\cos(\theta-\varphi)}\,d\omega \tag{4.10}
$$
$$
= -i\int_{|\omega|>|\mu|}\mathrm{sgn}(\omega)\tilde{p}(\theta,\omega,\mu)e^{i\omega r\cos(\theta-\varphi)}\,d\omega
$$

Substitute (4.10) in (4.1) and perform the Fourier series expansion with respect to $\varphi$, we have

$$
\int_0^{2\pi} e^{-in\varphi}\,d\varphi \int_0^{2\pi} e^{\mu r\sin(\theta-\varphi)}\int_{|\omega|>|\mu|}\mathrm{sgn}(\omega)\tilde{p}(\theta,\omega,\mu)e^{i\omega r\cos(\theta-\varphi)}\,d\omega\,d\theta = 0. \tag{4.11}
$$

Equation (4.11) is equivalent to

$$
\int_{|\omega|>|\mu|}\left[\int_0^{2\pi} e^{\mu r\sin\varphi+i(n\varphi+\omega r\cos\varphi)}\,d\varphi\right]\mathrm{sgn}(\omega)\tilde{p}_n(\omega,\mu)\,d\omega = 0. \tag{4.12}
$$

We cite one relation from the Appendix in [8] as follows

$$
\frac{1}{2\pi}\int_0^{2\pi} e^{a\sin\theta+i(n\theta+b\cos\theta)}\,d\theta = i^n\left(\frac{a+b}{\sqrt{b^2-a^2}}\right)^n J_n(\sqrt{b^2-a^2}), \quad |b|>|a|. \tag{4.13}
$$

With (4.13), (4.12) becomes

$$
\int_{|\mu|}^{\infty}\left[\left(\frac{\mu+\omega}{\sqrt{\omega^2-\mu^2}}\right)^n\tilde{p}_n(\omega,\mu)-\left(\frac{\mu-\omega}{\sqrt{\omega^2-\mu^2}}\right)^n\tilde{p}_n(-\omega,\mu)\right]J_n(r\sqrt{\omega^2-\mu^2})\,d\omega = 0, \tag{4.14}
$$

where $J_n(\omega)$ is the $n$th-order Bessel function. For more details of the Bessel functions and Hankel transform, we refer to [34]. Changing variable $\omega = \sqrt{\rho^2+\mu^2}$, (4.14) becomes

$$
\int_0^{\infty}\left(\frac{\mu+\sqrt{\rho^2+\mu^2}}{\rho}\right)^n\tilde{p}_n(\sqrt{\rho^2+\mu^2},\mu)\frac{1}{\sqrt{\rho^2+\mu^2}}J_n(r\rho)\rho\,d\rho
$$
$$
= \int_0^{\infty}\left(\frac{\mu-\sqrt{\rho^2+\mu^2}}{\rho}\right)^n\tilde{p}_n(-\sqrt{\rho^2+\mu^2},\mu)\frac{1}{\sqrt{\rho^2+\mu^2}}J_n(r\rho)\rho\,d\rho. \tag{4.15}
$$

Applying the $n$th-order Hankel transform to (4.15) with respect to variable $r$, we have

$$
\left(\frac{\mu+\sqrt{\rho^2+\mu^2}}{\rho}\right)^n\tilde{p}_n(\sqrt{\rho^2+\mu^2},\mu) = \left(\frac{\mu-\sqrt{\rho^2+\mu^2}}{\rho}\right)^n\tilde{p}_n(-\sqrt{\rho^2+\mu^2},\mu). \tag{4.16}
$$

Let $\sqrt{\rho^2+\mu^2} = \omega$ for $|\omega|>|\mu|$, (4.16) becomes

$$
(\mu+\omega)^n\tilde{p}_n(\omega,\mu) = (\mu-\omega)^n\tilde{p}_n(-\omega,\mu). \tag{4.17}
$$

This completes the proof. $\square$

# V.  Conclusion and Discussion

With the help of Plemelj formula, we have derived several explicit inversion formulas (2.1), (2.24) and (2.25) based on the inversion formulas of [4] for the complex constant $\mu$. Formula (2.1) can be regarded as the circular harmonic expression of two FBP formulas in [4]. Formula (2.25) is in the expression of exterior formula and may be used to reconstruct the function if only exterior data is available. The exterior formula (3.13) was first proved by Puro for pure imaginary $\mu$ and in this paper it is proved to hold for complex constant $\mu$ and the radius-dependent complex attenuation $\eta(r)$. The formula (3.13) $\mu \in R^1$ did not appear in the literature from our knowledge.

There are four different range conditions. In Section IV, we analyzed the relations among multiple range conditions and then used the circular harmonic expansion to show that Novikov's range condition (4.1) yields the traditional range condition (4.9) if $\mu \in R^1$. It is desired to further investigate the relations among these range conditions.

## Appendix. Properties of the Plemelj formulas

Assume that $\varphi(\varsigma)$ is a Hölder continuous function on $S^1$. We define $\Phi(z)$ as the Cauchy integral

$$\Phi(z) = \frac{1}{2\pi i} \int_{S^1} \frac{\varphi(\varsigma)}{\varsigma - z} d\varsigma .$$ (A-1)

It is easy to see that $\Phi(z)$ is analytical in $C \setminus S^1$. Denote by $\Phi^+(z)$ and $\Phi^-(z)$ the restrictions of $\Phi(z)$ inside and outside of $S^1$, respectively. On $S^1$, (A-1) is understood as the principal integral. We expand $\varphi(\varsigma)$ in the Fourier series

$$\varphi(\varsigma) = \sum_{-\infty}^{\infty} a_n \varsigma^n , \ \varsigma \in S^1 .$$ (A-2)

Three properties of the Plemelj formulas are summarized below:

$$\Phi^+(z) = \sum_{n \geq 0}^{\infty} a_n z^n , \ |z| < 1 .$$ (A-3)

$$\Phi^-(z) = \sum_{n < 0}^{\infty} a_n z^n , \ |z| > 1 .$$ (A-4)

$$\frac{1}{2\pi i} \int_{S^1} \frac{\varphi(\varsigma)}{\varsigma - \varsigma_0} d\varsigma = \frac{1}{2} (\sum_{n \geq 0}^{\infty} a_n \varsigma_0^n - \sum_{n < 0}^{\infty} a_n \varsigma_0^n) , \ \varsigma_0 \in S^1 .$$ (A-5)

For the details on the derivation of (A-3), (A-4) and (A-5) and other properties about the Plemelj formulas, we refer to [27]. Let $f(z)$ be analytical in a region that contains $\overline{D}$, for $z_0 \in S^1$, an immediate result from (A-5) is

$$\frac{1}{2\pi i} \int_{S^1} \frac{f(\varsigma)}{\varsigma - z_0} d\varsigma = \frac{1}{2} f(z_0) .$$ (A-6)